# Interpretation of relativistic, transverse, and longitudinal mass using the Lorentz transformation of reference time: Explanation of time dilation via spherical light clock

**Masanori Sato**

*Honda Electronics Co., Ltd.,*

*20 Oyamazuka, Oiwa-cho, Toyohashi, Aichi 441-3193, Japan*

**Abstract**: An interpretation of the inertial mass increase due to an object's velocity which is derived from the theory of special relativity is discussed. A Lorentz transformation of the reference time causes the inertial mass increase. It is assumed that the real physical phenomenon that occurs is an expansion of the reference time by a Lorentz transformation (that is, a dilation of time). Only the reference time is assumed to be variant, which causes relativistic mass. Time dilation is real. Relativistic mass originates in a Lorentz transformation of reference time.

1. Introduction

In this discussion, the speed of light $c$ is assumed to be invariant. The physical reality is only time dilation by velocity. The constancy of the speed of light causes the time dilation by motion. Mass m and length $x$ are also invariant though mass and length appear to be variant through the Lorentz transformation of reference time. That is, Lorentz contraction of length and inertial mass increase can also be explained by the Lorentz transformation of reference time. Length $x$ appears to be variant in the geometric properties of space-time. Time dilation by the velocity is explained using the constancy of the speed of light $c$. An inertial mass is explained by the time dilation. Moreover, a light clock [1] shows that in a moving frame a photon travels a longer path than in a stationary state. This fact causes not only time dilation but also inertial mass, that is, transverse and longitudinal mass.

Sachs [2] described that the Lorentz-Fitzgerald contraction is not physical change but scale change: "But this is nothing more than a *scale change* in the expression of the physical laws in the respective frames of reference. It does not at all refer to a physical change of a material body, such as the shortening of a meter stick that is in motion, by virtue of its motion relative to an observer. The latter 'physical change' would require dynamical laws of matter for their prediction. The scale change, on the other hand, is only a kinematic relation." He also described that the time contraction is *scale change*: "this does not signify a physical change of duration in one frame of reference compared with the other." and "This is not more than a *scale change* of the measure of duration in the moving frame compared with the frame of the observer's

clock."

Taylor and Wheeler [3] argued against the term "relativistic mass" in their book which states, "In reality, the increase of energy with velocity originates not in the object but in the geometric properties of space-time itself." Okun [4] described Newtonian mass m, which does not vary with velocity: "These days, when physicists talk about mass in their research, they always mean invariant mass." Oas [5, 6] discussed and summarized the historical use of the concept of relativistic mass. He reviewed more than 400 books on special and general relativity, classifying as to whether or not they introduced the concept of relativistic mass or not. He showed that around 2/3 introduced the concept of relativistic mass, but that regarding introductory or modern physics text books, published recently, the number of books that do not introduce the concept of relativistic mass was large.

The Michelson-Morley interference experiment does not measure the flight time of photons traveling in two paths. This is because interference appears in single photon interference experiment. As discussed in the previous report [7], a single photon Michelson-Morley type experiment can be carried out. That is, there is a single photon in the Michelson-Morley experimental setup and the interference can be observed. I have discussed this problem using the de-Broglie Bohm picture. At this stage, I think the hypothesis of frame dragging of the permittivity $\varepsilon_0$ and permeability $\mu_0$ by the gravitational field of the earth is suitable [8, 9]. That is, like sound in the atmosphere around the earth, light travels in the permittivity $\varepsilon_0$ and permeability $\mu_0$ in the gravitational field of the earth. Sound is not affected by the earth's motion: this is because the atmosphere is dragged by the earth. The classic hypothesis of frame dragging (of the permittivity $\varepsilon_0$ and permeability $\mu_0$) can explain the Michelson-Morley experiment. The gravitational field of the earth is largely considered as a stationary state.

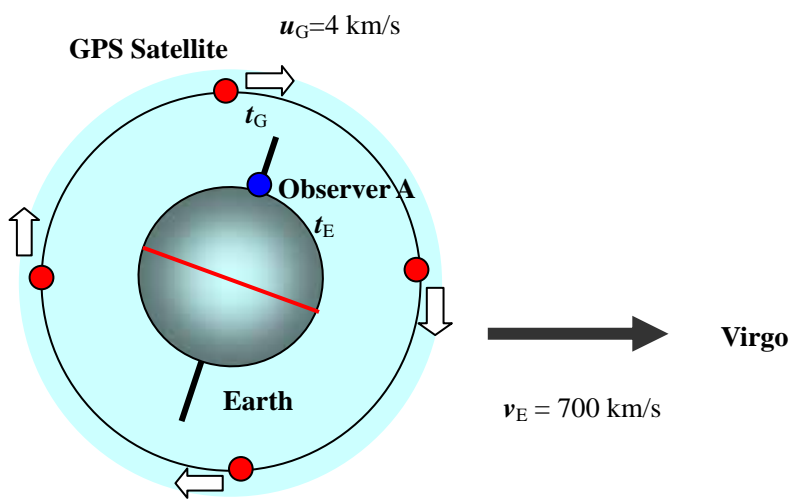

**Fig. 1** Illustration of the stationary and moving frames in the CMB: The stationary state is on the north pole of the earth and the moving frame is the GPS satellite. The velocity of the earth in the CMB $v_E$=700 km/s does not affect the earth or the GPS satellite. The velocity of the GPS satellite $u_G$ =4 km/s observed in the ECI coordinate system can be assumed to be the absolute velocity. The time dilation of the GPS satellite is 7.1 μs every day.

The stationary state and moving frame are defined. These experiments are carried out in the gravitational field of the earth. **Figure 1** shows an illustration: the stationary state is at the north pole of the earth and the moving frame is the global positioning system (GPS) satellite. For a more simplistic example, consider observer A is located at the North Pole, eliminating the earth's rotation, and neglecting the effect of the gravitational potential of the earth. The gravitational field of the earth can be provisionally considered to be the stationary state [8, 9]. A velocity of 700 km/s in the cosmic microwave background (CMB) does not affect the GPS satellite or the observer on earth. Thus, the velocity of the GPS satellite $u_G$ =4 km/s observed in the earth-centered locally inertial (ECI) coordinate system can be assumed to be an absolute velocity. Observer A on earth sees a time dilation on the clock in the GPS satellite, however, the observer in the GPS satellite sees a time gain on the clock on earth. This is because only the reference time in the GPS satellite becomes large.

Where, $t_E$ is the reference time on earth, $t_G$ is that of in the GPS satellite. If the effect of the gravitational potential of the earth is eliminated, it is assumed that $t_E = t_0$, ($t_0$ is the reference time in the absolute stationary state).

$$t_G = \frac{t_E}{\sqrt{1-\left(\frac{4}{300000}\right)^2}}$$

The reference time expansion according to Lorentz transformation denotes that the phenomena progress slowly when seen from a stationary state. In **Fig. 1**, the time dilation by velocity only occurs on the GPS satellite. Observer A does not suffer the time dilation by the velocity. This is because observer A is provisionally in a stationary state (the effect of the gravitational potential is neglected.). The GPS satellite orbits the earth, and thus, only a transverse Doppler shift is detected. The difference is calculated as follows,

$$\frac{t_E - t_G}{t_E} = 1 - \frac{1}{\sqrt{1-\left(\frac{4}{300000}\right)^2}} = -0.889 \times 10^{-10}.$$

The time dilation of the GPS satellite is $0.889 \times 10^{-10} \times 60 \times 60 \times 24 = 7.1 \mu s$ every day [8].

Lorentz contraction is also discussed. Car navigation systems using the GPS precisely work [10]. If the stationary state is assumed to be the CMB, the absolute velocity of the earth is estimated to be 700 km/s as shown in **Fig. 2**. Thus, the Lorentz contraction is calculated as

$$1 - \frac{1}{\sqrt{1-\left(\frac{700}{300000}\right)^2}} = -2.7 \times 10^{-6}.$$

Therefore, the deviation of a car navigation system is maximally estimated as 54 m every 12 hours [11]. The deviation of the radius of the earth is estimated to be 17.5 m. Of course, there are no such deviations detected; that is, the car navigation system appears to work precisely [10]. This indicates that Lorentz contraction originates not in the object but in the geometric properties of space-time itself (that is, not physical change but scale change): we cannot detect any Lorentz contraction in the gravitational field of the earth. In GPS experiments, only time dilation due to the velocity in the ECI coordinate system is clearly observed. According to the GPS experimental results, this is one way to directly measure distance using the principle of the isotropic constancy of the speed of light - there is no Lorentz contraction detected. This is the reason that length *x* is assumed to be invariant. In this report, the physical meaning of relativistic effect is explained using only the Lorentz transformation of reference time.

Velocity $u = \dfrac{dx}{dt}$ contains time by differential form, the numerator of this form is assumed to be invariant, but the denominator is variant according to velocity, therefore velocity *u* is variant. The relativistic velocity addition law and inertial mass increase due to the velocity are explained by a Lorentz transformation of reference time.

The transverse mass is defined by electromagnetic force (Lorentz force) and the longitudinal mass is defined by electrostatic force.

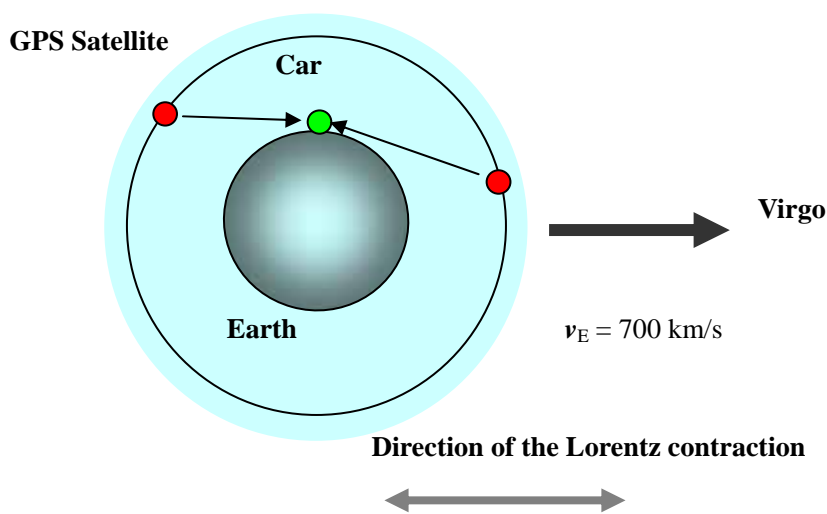

**Fig. 2** Illustration of the earth motion in the CMB. The earth motion is assumed to be at some 700 km/s towards the constellation Virgo. At a moment when a car on earth faces two GPS satellites, the car detects the speed of light *c* from two GPS satellites. According to the theory of special relativity, the Lorentz contraction of $2.7 \times 10^{-6}$ occurs in the direction of $v_E$, when seen from the CMB. However, there is no Lorentz contraction detected in the GPS. This is because Lorentz contraction originates in the geometric properties of space-time itself:

2. Summary of the assumptions of this proposal

In this section, I will summarize the difference between the orthodox interpretation and this proposal. Table 1 shows the critical differences. I do not consider that the theory of special relativity can be applied to the experiments in the gravitational field of the earth. That is, Minkowski space (free space) cannot apply to the gravitational field of the earth, so the theory of general relativity should be adopted.

Absolute stationary state is a very convenient concept. I consider that the gravitational field of the earth is substantially the stationary state. That is, the gravitational field of the earth is not affected by the motion of the earth. This is confirmed by the results of the GPS experiment where there is no Lorentz contraction of length in the gravitational field of the earth. Here, I provisionally consider that the contraction of length with velocity originates not in the object but in the geometric properties of time itself.

Table 1 Critical difference

|   | Terms | Orthodox interpretation | This proposal |
|---|---|---|---|
| 1 | Absolute stationary state | No | Yes |
| 2 | Lorentz contraction of length in the gravitational field of the earth | Yes | No |
| 3 | Can the theory of special relativity be applied to the experiment in the gravitational field of the earth? | Yes | No |
| 4 | Hypothesis of frame dragging of the permittivity $\varepsilon_0$ and permeability $\mu_0$ by the gravitational field of the earth | No | Yes |

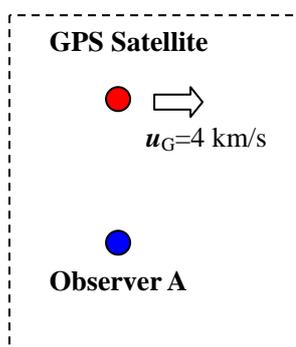

**Fig. 3** Illustration that shows $u_G$=4 km/s is not a relative velocity in the gravitational field of the earth. This is because the reference time of GPS satellite expands by Lorentz transformation, which was confirmed experimentally. Therefore, when the GPS satellite sees observer A, the GPS satellite obtains a smaller velocity than $u_G$=4 km/s.

Table 2 Observation of GPS experiments

|   | Observation of GPS experiments | Experimental results |
|---|---|---|
| 1 | Observer A sees GPS satellite | Time dilation |
| 2 | GPS satellite sees observer A | Time gain |

**Figure 3** and Table 2 show that the velocity $u_G$=4 km/s is not a relative between observer A and the GPS satellite. This is because the reference time of the GPS satellite expands by Lorentz transformation, which was confirmed experimentally as time dilation on the GPS satellite. Experimental results are: Observer A sees GPS satellite - time dilation, GPS satellite sees observer A - time gain. Therefore, when the GPS satellite sees observer A the GPS satellite obtains a velocity smaller than $u_G$=4 km/s. Let us represent that velocity as $u_{G-A}$, where the subscript G-A represents that the GPS satellite sees observer A.

$$\frac{u_{G-A}}{u_G} = \frac{t_E}{t_G} = \sqrt{1 - \left(\frac{4}{300000}\right)^2} = 0.99999999991.$$

3. Absolute stationary state

In the discussion of the theory of special relativity, the absolute stationary state has always been denied. However, I consider the absolute stationary state to be a very convenient concept. The most familiar absolute stationary state is the gravitational field of the earth, if a gravitational potential is avoided. If the absolute stationary state is assumed, there arises no twin paradox. For example, let us assume that the absolute velocity is 30 km/s. The effects of acceleration and deceleration are negligible. The effect of the Lorentz factor is calculated as follows,

$$\frac{t_0 - t_u}{t_0} = 1 - \frac{1}{\sqrt{1-\left(\frac{30}{300000}\right)^2}} = -0.500 \times 10^{-8}.$$

Thus, the twin grows younger every day as follows,

$$\therefore 0.500 \times 10^{-8} \times 24 \times 60 \times 60 = 432 \times 10^{-6} = 432\,\mu s.$$

If we neglect the difference in gravitational potential of the earth, the traveling twin is 432 μs younger every day.

According to this hypothesis, if the CMB is the absolute stationary state, the rocket suffers an absolute velocity of 700 km when the rocket leaves the gravitational field of the earth. Thus the twin in the rocket is $2.7 \times 10^{-6} \times 60 \times 60 \times 24\,s = 0.23\,s$ younger every day than the twin on earth. This time dilation is large enough to detect; for example, the spacecraft Galileo may possibly detect a time dilation of $0.23 \times 365 = 84\,s$ /year.

At this stage, I assume a hypothesis of the dragging of the permittivity $\varepsilon_0$ and permeability $\mu_0$, however, I do not know where the spacecraft suffers the drift of the permittivity $\varepsilon_0$ and permeability $\mu_0$. At least, the GPS satellites, which orbit 20,000 km above the ground level, do not suffer the drift of the permittivity $\varepsilon_0$ and permeability $\mu_0$.

4. Spherical light clock in motion and the Lorentz transformation

In this section, time dilation by the velocity is explained using the constancy of the speed of light $c$, thereafter an inertial mass is explained by the time dilation. A light clock [1] shows that in a moving frame a photon travels a longer path than in a stationary state. This causes not only time dilation but also inertial mass (that is, transverse and longitudinal mass).

Feynman [1] used a linear light clock to visualize time dilation by motion. As discussed in the previous paper [12], the Lorentz transformation of reference time is derived using a spherical light clock, in which isotropic motion can be explained. Discussion concerning the direction of the light clock motion is avoidable. To obtain the physical meanings of the Lorentz transformation, let us consider a spherical light clock, which is isotropic for motion, shown in **Fig. 4**. The original light clock has the problem of direction of motion. A photon that is radiated from photon source P is reflected by the spherical shell and back to the photon source P. At the stationary state, the reference time $t_0$ is defined as follows,

$$t_0 = \frac{2L}{c} , \tag{1}$$

where L is the radius of a spherical shell.

**Figure 5** shows the light clock in motion at velocity *u*. In this condition, the Pythagorean theorem can be applied.

The speed of light is assumed to be constant and independent on the motion of the light source. When the system moves at velocity *u*, point A moves to point A': thus a photon has to move the distance OA' (the traveling time of the distance OA' is represented as $t_u$). From the Pythagorean theorem we obtain, as follows,

$$\left(\frac{t_u c}{2}\right)^2 = L^2 + \left(\frac{t_u u}{2}\right)^2 = \left(\frac{t_0 c}{2}\right)^2 + \left(\frac{t_u u}{2}\right)^2 . \tag{2}$$

$$\therefore t_u^2 = \frac{(t_0 c)^2}{(c^2 - u^2)} , \tag{3}$$

where, the subscripts 0 and u represent the reference frame at rest and the moving frame at velocity *u*, respectively ($t_0$ is the reference time in the stationary state, and $t_u$ is that of the moving frame at velocity *u*). The Lorentz transformation of the reference time moving at velocity *u* is represented as equation (4).

$$t_u = \frac{t_0}{\sqrt{1 - \left(\frac{u}{c}\right)^2}} , \tag{4}$$

Equation (4) defines the reference time of the moving clock. At the same time equation (4) shows the Lorentz transformation of reference time. Equation (4) is rewritten using a differential form as equation (5).

$$\therefore \frac{dt_u}{dt_0} = \frac{1}{\sqrt{1 - \left(\frac{u}{c}\right)^2}} . \tag{5}$$

Equation (5) shows Lorentz factor.

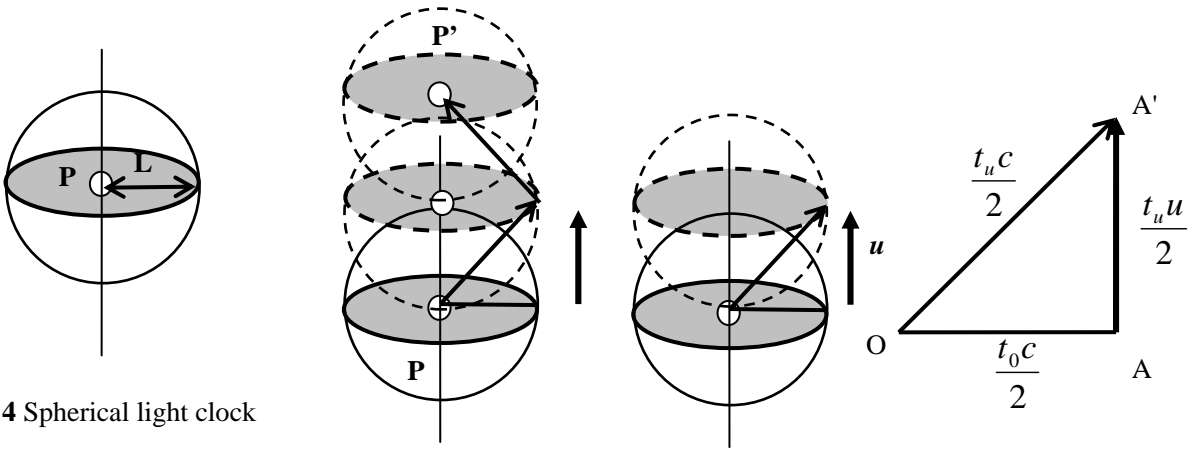

**Fig. 4** Spherical light clock

**Fig. 5** Spherical light clock in motion

**Figure 5** shows that the spherical light clock is isotropic in motion. That is, the direction of spherical light clock in motion (velocity $u$) is always orthogonal to the plane of photons that take part in the light clock (gray circular plane). The moving direction of the light clock is orthogonal to that of a photon. The direction of the light clock motion (represented as vector $u$) is always orthogonal to that of a photon path in the light clock. That is, the light clock is always orientated orthogonal to the direction of velocity $u$, as shown in **Fig. 6**.

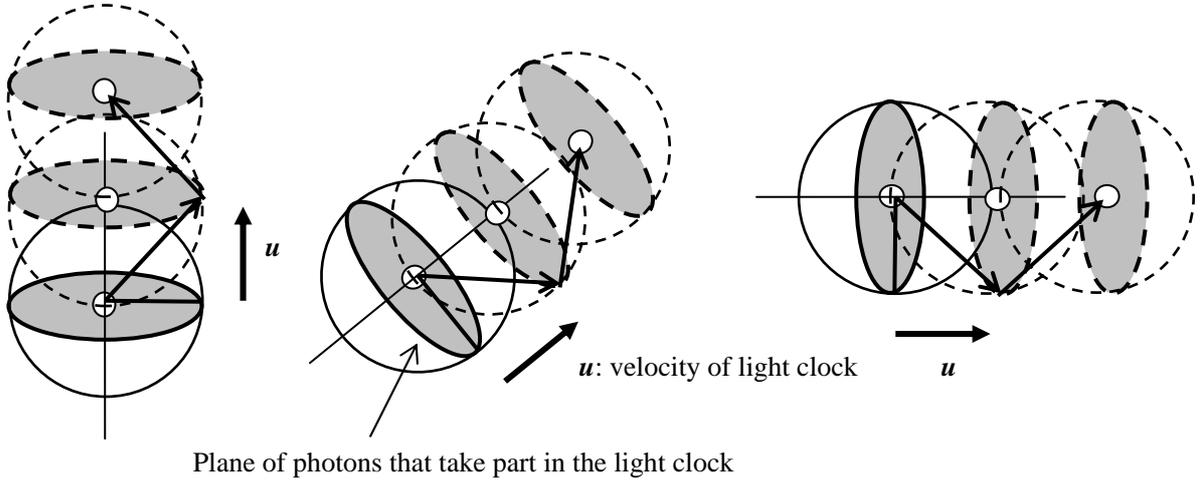

Plane of photons that take part in the light clock

**Fig. 6** Spherical light clock in motion: for all directions of motion, the plane of photons of the light clock is always orthogonal to velocity $u$. Therefore, the time dilation of spherical light clock is isotropic.

5. Application of the Lorentz transformation to the relativistic velocity addition law

Velocity u is related to the reference time by a differential equation, therefore velocity u is the variant. Taylor and Wheeler [1] described that the increase of momentum with velocity originates in the geometric properties of space-time itself. However, I consider that the increase of momentum with velocity originates in the Lorentz transformation of time.

Let us consider two frames. One is in the stationary state and the other is a moving frame at velocity $u$. When the phenomena that occurred in the moving frame are seen from the stationary state, they are seen to proceed slowly. This is because the reference time becomes large according to equation (4). In these discussions, the representation $|_O$ indicates that the quantity is evaluated in the stationary state and $|_U$ indicates that it is evaluated in the frame moving with velocity $u$. I consider the addition of velocity $\Delta v_0$ in the frame moving at velocity $u$ where $\Delta v_0$ is a small velocity in the stationary state.

$$\Delta v_0 = \left.\frac{\Delta x}{\Delta t}\right|_O \equiv \frac{dx}{dt_0} \quad .$$

**Figure 7** shows the small velocity $\Delta v_u$ in the moving frame as seen from the stationary state. $\Delta v_u$ is defined as follows,

$$\Delta v_u = \left.\frac{\Delta x}{\Delta t}\right|_U \equiv \frac{dx}{dt_u}$$

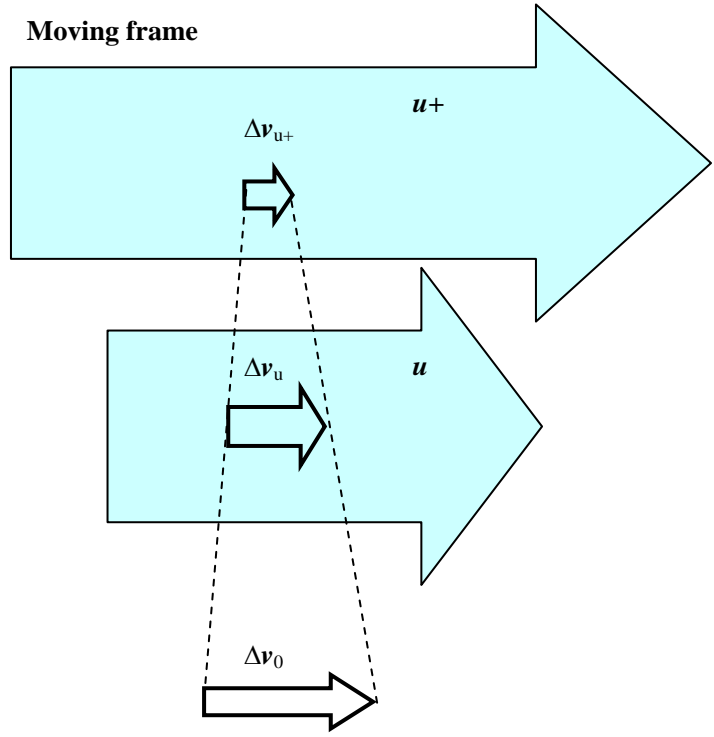

**Fig. 7** Velocity contraction (velocity in the moving frame as seen from the stationary state): $\Delta v_0$ is the small velocity in the stationary state, $u$ is the velocity of the moving frame in the stationary state, and $\Delta v_u$ is the small velocity in the moving frame seen from the stationary state where, the velocities $\Delta v_0$ and $\Delta v_u$ are combined by the Lorentz transformation. The vector $\Delta v_u$ contracts according to the velocity $u$. Where, $u+$ is a larger velocity.

Lorentz contraction is not assumed; therefore, $x$ has the same value in the moving frame, that is, $\Delta x|_O = \Delta x|_U = dx$. The differential reference times are $\Delta t|_O = dt_0$ and $\Delta t|_U = dt_u$. Only the differential reference time of the differential in the moving frame is enlarged according to equation (5); thus, equation (6) is obtained.

$$\frac{dx}{dt_u} = \frac{dx}{dt_0}\frac{dt_0}{dt_u} = \Delta v_0 \sqrt{1-\left(\frac{u}{c}\right)^2} = \Delta v_u \quad . \tag{6}$$

Equation (6) shows the velocity contraction by the Lorentz transformation. Equation (6) is rewritten as follows,

$$\Delta v_0 = \frac{dx}{dt_0} = \frac{dx}{dt_u}\frac{dt_u}{dt_0} = \frac{1}{\sqrt{1-\left(\frac{u}{c}\right)^2}}\frac{dx}{dt_u} = \frac{\Delta v_u}{\sqrt{1-\left(\frac{u}{c}\right)^2}} \quad . \tag{7}$$

Equations (6) and (7) show that, in a reference frame moving at velocity $u$, the reference time becomes large according to equation (4). Thus, the addition of $\Delta v_0$ in the moving frame seen from the stationary state is represented as,

$$\therefore u + \Delta v_u = u + \sqrt{1 - \left(\frac{u}{c}\right)^2} \Delta v_0 \quad . \tag{8}$$

Equation (8) is the velocity addition law assuming the Lorentz transformation of the reference time.

6. Interpretation of inertial mass increase

The inertial mass increase due to the velocity is observed in the deflection in the magnetic field. The momentum *P'* of a moving particle at velocity *u* is represented by equation (9) where m' is the relativistic mass at velocity *u* and m is the inertial mass in the stationary state.

$$p' = m'u = \frac{m}{\sqrt{1 - \left(\frac{u}{c}\right)^2}} u \quad . \tag{9}$$

Equation (9) shows that the inertial mass m' of the moving particle is increased by virtue of having a relative velocity. Let us try to explain equation (9) using the Lorentz transformation of reference time. That is, using equations (4) and (5), only the reference time's expansion is assumed. The inertial mass and length are not modified; that is, $m|_U = m$, $x|_U = x$. The physical meaning is as follows. In the moving frame, the reference time is expanded. Thus, the effect of phenomena on moving particles seen from the stationary state occurs more slowly. The mechanism of time dilation acts on the electromagnetic phenomena. A photon that transfers the electromagnetic force has to travel a longer path in the moving frame.

Using the Lorentz transformation in equation (5), equation (9) is rewritten as follows,

$$p' = m'u = m' \frac{dx}{dt}\bigg|_O = \frac{m}{\sqrt{1 - \left(\frac{u}{c}\right)^2}} \frac{dx}{dt}\bigg|_O = m \frac{dt_u}{dt_0} \frac{dx}{dt}\bigg|_O \quad , \tag{10}$$

where, the differential $\frac{dx}{dt}\bigg|_O$ shows the velocity in the stationary state, and the coefficient $\frac{dt_u}{dt_0}$ denotes the Lorentz factor. That is, in the moving frame at velocity *u*, the reference time expands according to equation (5) and the phenomena progress slowly. Thus, we see the phenomena in the moving frame from the stationary state using equation (9), which are rewritten using the coefficient $\frac{dt_u}{dt_0}$ as shown in equation (10).

Let us consider reference time and progress speed of the phenomena in the moving frame. When we see velocity $\Delta v_0$ in the moving frame from the stationary state, we obtain the value of $\Delta v_u$. That is, when we see the velocity in the moving frame from stationary state, it looks small according to equation (6). The phenomena in the moving frame seen from the stationary state seems to proceed slower by the coefficient

$$\left(\frac{dt_u}{dt_0}\right)^{-1} = \frac{dt_0}{dt_u} = \sqrt{1-\left(\frac{u}{c}\right)^2}. \tag{11}$$

7. Discussion

7.1 Comparison of velocity addition

There is a tendency for the value of added velocity to become smaller than simple summation. If $u \sim c$, there is no increase in velocity. Thus, velocity $c$ is the upper limit. Let us compare the calculated values using equation (8) and the relativistic velocity addition law [13]. Table 3 shows the comparison between the three equations. Calculations are carried out on the condition $u=0.75c$, $\Delta v=0.1c$. When $\Delta v \langle\langle c$, there is a little difference between equations (8) and (12).

Table 3 Comparison of velocity addition

|  | Addition of $\Delta v$ on $u$ | $u=0.75c$, $\Delta v=0.1c$ | Upper limit |
|---|---|---|---|
| Equation (8) | $u + \sqrt{1-\left(\frac{u}{c}\right)^2}\Delta v$ (8) | $0.8161c$ | $c$ |
| Relativistic velocity addition Law [13] | $\dfrac{u + \Delta v}{1 + \dfrac{u \cdot \Delta v}{c^2}}$ (12) | $0.7906c$ | $c$ |

7.2 Physical meanings of the velocity $u$ and time compensation

This section discusses the physical meanings of the velocity $u$. According to equation (8), the procedure of acceleration to obtain velocity $u$ (that is, the additions of the small velocities) contains an effect of the Lorentz transformation of reference time. Equation (8) shows that the acceleration should be represented using an integral.

Let us consider time compensation by velocity $u$. **Figure 8** shows that the multiplication of the coefficient of Lorentz factor, $\dfrac{dt_u}{dt_0} = \dfrac{1}{\sqrt{1-\left(\dfrac{u}{c}\right)^2}}$ shows time compensation which modifies the phenomena in the moving frame as seen from the stationary state. Time progresses slowly in the moving frame when the phenomena are seen from the stationary state. In the moving frame, it takes much more time for the photon to travel as discussed in the moving light clock; that is, the reference time expansion represented as the Lorentz transformation, or the coefficient $\dfrac{dt_u}{dt_0}$ multiplies. The momentum increase is experimentally observed. In this report, the momentum increase is not interpreted by the inertial mass increase but by the reference time expansion. The physical reality is only an expansion of reference time.

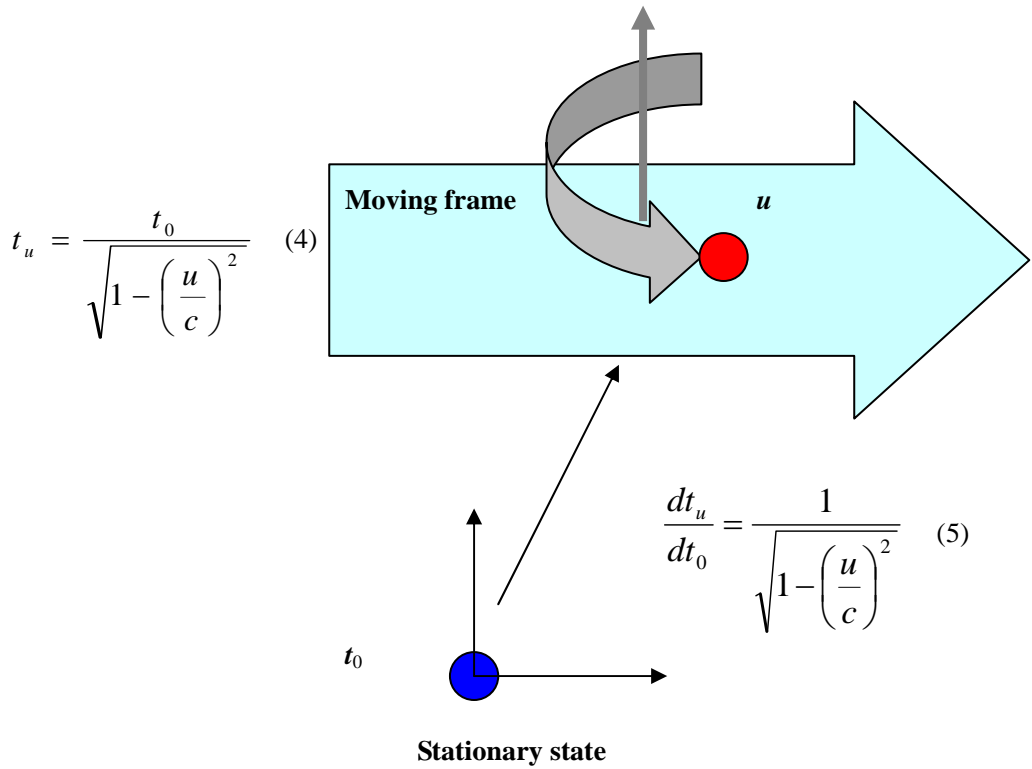

$$t_u = \frac{t_0}{\sqrt{1-\left(\frac{u}{c}\right)^2}} \quad (4)$$

$$\frac{dt_u}{dt_0} = \frac{1}{\sqrt{1-\left(\frac{u}{c}\right)^2}} \quad (5)$$

**Fig. 8** Time compensation: When the phenomena in the moving frame are seen from the stationary state, a time compensation is required. The time compensation is carried out using the Lorentz transformation of reference time. That is, the reference time expansion in the moving frame is compensated by equation (5). This is because phenomena progressing slowly are equivalent to the inertial mass increase; however, this should be interpreted as a momentum increase. These phenomena (for example, the inertial mass increases in Lorentz force) are experimentally observed. The moving particle looks to have a larger inertial mass, but this should be interpreted as the Lorentz transformation of reference time.

I consider the explanation that the phenomena progress slowly in the moving frame to be acceptable, however, it is rather difficult to accept the physical reality of the momentum increase. Physically, the coefficient $\frac{dt_u}{dt_0}$ in equation (10) is related to the velocity *u* not the inertial mass. This is because the velocity addition represented in equation (8), which is the multiplication of $\left(\frac{dt_u}{dt_0}\right)^{-1} = \sqrt{1-\left(\frac{u}{c}\right)^2}$ already contains the effect of the Lorentz transformation of reference time, and therefore compensation by the coefficient $\frac{dt_u}{dt_0}$ is required.

7. 3 Transverse mass

There have been discussions of the transverse and longitudinal mass [1, 2]. The mass moving in the transverse direction with velocity *u* is called the transverse mass and that in the parallel direction is the longitudinal mass.

Let us consider a Lorentz force. When photons are in resonance at velocity *u*, the motion of photons that take part in the transverse phenomena is similar to a light clock in motion as shown in **Fig. 9**. That is, the photons in the gray plane dominantly affect the phenomena. This is because photon motion, which composes the Lorentz force, is transverse to the direction of the velocity *u*. The lorentz factor of equation (5) multiplies, and thus transverse mass $m_T$ is represented as,

$$m_T = m \times Lorentz\ factor = \frac{m}{\sqrt{1-\left(\frac{u}{c}\right)^2}}.$$

In this discussion, the Lorentz force is assumed. The direction of velocity *u* is transverse to the direction of force. Thus, the model of the moving light clock in **Fig. 4** seems to work well. That is, the direction of the photon motion, which transfers electromagnetic force, is transverse to that of the light clock motion. The traveling path length of a photon can be calculated using the Pythagorean theorem. For example, if *u*=0.9c in the transverse condition, the traveling path length is 2.3 times greater.

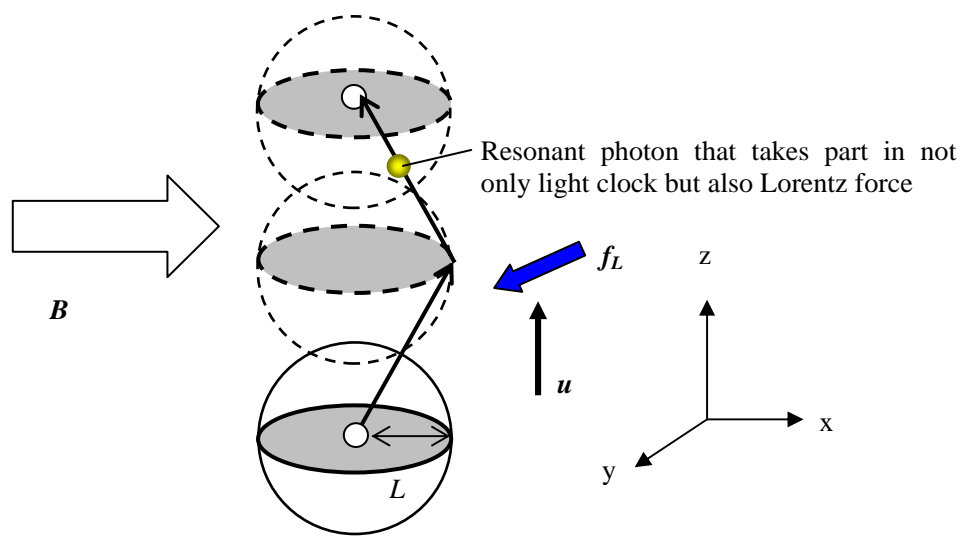

**Fig. 9** Transverse mass: The spherical light clock motion u is perpendicular to the Lorentz force $f_L$. A photon that takes part in the Lorentz force and the light clock, where $f_L \perp u$. *B* is magnetic field.

7.4 Longitudinal mass

In the condition that the direction of motion is parallel to that of force, not only the Lorentz factor but also the modification of photon path length should be applied. **Figure 10** shows electrostatic force where force direction is parallel to the velocity *u*. The round traveling time of photon $t_L$ is calculated as follows,

$$t_L = \frac{L}{c-u} + \frac{L}{c+u} = \frac{t_0}{1-\left(\frac{u}{c}\right)^2}. \tag{13}$$

The expansion represented in equation (13) indicates an inertial mass increase. This is because the electromagnetic or electrostatic force is transferred through photons, and equation (13) indicates that the force is weakened. However, in the representation of the longitudinal mass $m_L$ the Lorentz factor multiplies as follows,

$$m_L = \frac{m}{1-\left(\frac{u}{c}\right)^2} \times \frac{1}{\sqrt{1-\left(\frac{u}{c}\right)^2}}. \qquad (14)$$

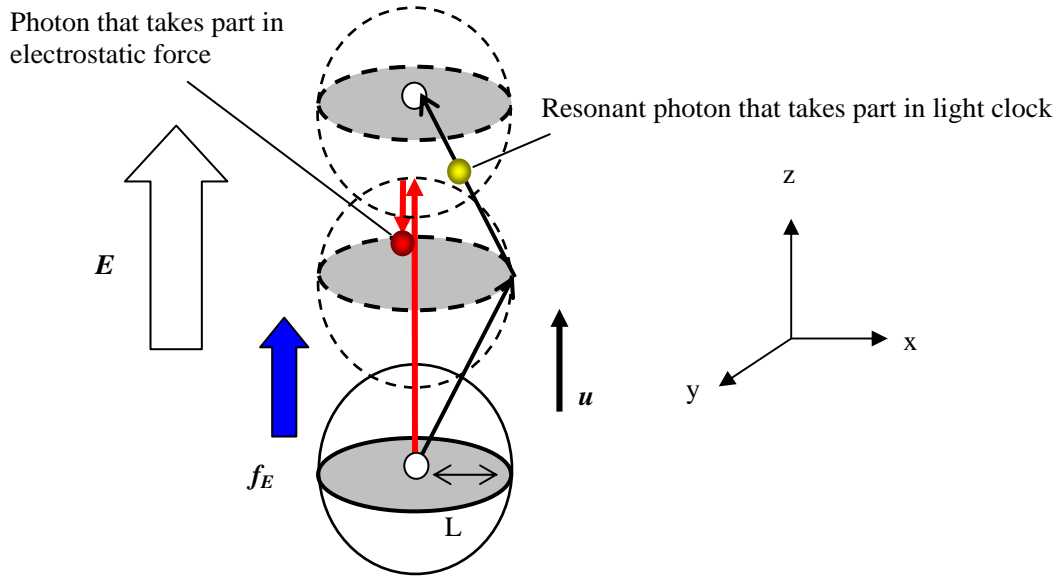

**Fig. 10** Longitudinal mass: Experimental condition is such that the spherical light clock moves parallel to electrostatic force $f_E$, that is, $f_E // u$. The path length of the photon that takes part in electrostatic force relates to the derivation of longitudinal mass. A photon which takes part in the electrostatic force is represented as a red circle. The time of the round trip of the photon (red circle) is represented in equation (13). $E$ is the electrostatic field.

The photon path of electrostatic force is represented by the red arrows. The calculations of the traveling photon path length are shown in Table 4. The calculations of the longitudinal mass are also shown. Photon path length and the representation of transverse mass are in good agreement, however, that of longitudinal mass $m_L$ requires not only the expansion in equation (13) but also the Lorentz factor.

Table 4 Photon path length and inertial mass increase

|  | $u$=0.9c |  | $u$=0.99c |  |
|---|---|---|---|---|
|  | Path length | Mass | Path length | Mass |
| Transverse | 2.3 $=\frac{1}{\sqrt{1-0.9^2}}$ | 2.3 | 7.1 $=\frac{1}{\sqrt{1-0.99^2}}$ | 7.1 |
| Longitudinal | 5.2 $=\frac{1}{1-0.9^2}$ | 12.1 $=\frac{1}{(1-0.9^2)\sqrt{1-0.9^2}}$ | 50 $=\frac{1}{1-0.99^2}$ | 356 $=\frac{1}{(1-0.99^2)\sqrt{1-0.99^2}}$ |

## 7.5 Summary using the geometric properties of space-time

Let us assume the geometric properties of space-time to be the Lorentz factor represented by the coefficient $\dfrac{dt_u}{dt_0}$ represented in equation (5). In this discussion, space is assumed to be invariant, thus, only the geometric properties of time is taken into consideration using equation (5).

As far as momentum is concerned, the velocity *u* seen in the stationary state should be compensated as *u'* represented as follows,

$$u' = u \times Lorentz\ factor = u\frac{dt_u}{dt_0} \qquad (15)$$

We cannot detect *u'* itself. It only can be detected as the momentum *P'* in the stationary state,

$$P' = mu' = mu\frac{dt_u}{dt_0} = mu \times Lorentz\ factor = \frac{mu}{\sqrt{1-\left(\dfrac{u}{c}\right)^2}} \ . \qquad (16)$$

Equation (16) is the representation of the transverse momentum.

The longitudinal mass $m_L$ indicates the effect of electrostatic force in the moving frame. The traveling path of the photon is parallel to velocity *u*. That is, the photon action is weakened in proportion to the inversion of the photon path length. At this stage, the multiplication of equation (13) and the coefficient of the Lorentz factor is used, and thus equation (14) is derived as follows,

$$m_L = m \times \frac{photon\ path\ length}{2L} \times Lorentz\ factor = \frac{m}{1-\left(\dfrac{u}{c}\right)^2} \times \frac{1}{\sqrt{1-\left(\dfrac{u}{c}\right)^2}}.$$

Where, the photon path length is the photon round trip path in the moving light clock. I consider this representation shows geometric properties.

## 8. Conclusion

In this report, the physical meaning of the time dilation of the theory of special relativity was discussed. That is, the physical change of the relativistic effect caused by the velocity is the Lorentz transformation of reference time. The speed of light *c*, is invariant. Mass m, and length *x* are also assumed to be invariant, though they look variant via scale change. However, only the reference time is assumed to be variant. This report describes that the physical reality of the relativistic effects are caused by the Lorentz transformation of reference time.